\documentclass[12pt]{article}
\usepackage{amssymb}
\usepackage{mathtools}
\usepackage{hyperref}

\usepackage{fullpage}
\numberwithin{equation}{section}

\usepackage{titlesec}
\titlespacing*{\section}
{0pt}{0ex}{0ex}
\titlespacing*{\subsection}
{0pt}{0ex}{0ex}

\usepackage{natbib,graphicx,setspace}
\usepackage{amsmath,amsthm,bm,color}

\usepackage{xcolor}         
\usepackage{fullpage,amsmath}                        
\usepackage{caption}
\usepackage{hyperref}
\usepackage[rightbars, color]{changebar}
\usepackage{multirow}

\definecolor{DarkGreen}{rgb}{0.5,0.8,0.6}   
\definecolor{RGBblack}{rgb}{0.0,0.0,0.0}    

\def\hei{\color{black}}

\newcommand{\ech}{\hei \rm}

\def\stackover#1#2{\mathrel{\mathop{#2}\limits^{#1}}}
\newcommand{\iid}{\stackover{\mbox{\footnotesize i.i.d.}}{\sim}}

\newcommand{\is}{\itemsep=0pt}
\newcommand{\bd}[1]{\begin{description}[#1]\is}
  \newcommand{\ed}{\end{description}}
\newcommand{\bi}{\begin{itemize}\is}
  \newcommand{\ei}{\end{itemize}}
\newcommand{\be}{\begin{enumerate}\is}
  \newcommand{\ee}{\end{enumerate}}
  \newcommand{\beq}{\begin{eqnarray}\is}
  \newcommand{\eeq}{\end{eqnarray}}
\makeatletter
\newcommand*{\rom}[1]{\expandafter\@slowromancap\romannumeral #1@}

\newcommand{\bY}{\bm{Y}}

\newcommand{\Ms}{M^\star}
\newcommand{\tM}{\widetilde{M}}

\renewcommand{\th}{\theta}
\newcommand{\eps}{\epsilon}
\newcommand{\sig}{\sigma}

\newcommand{\etahat}{\widehat{\eta}}

\newcommand{\Ubar}{\overline{U}}

\newcommand{\bth}{\bm{\th}}

\newcommand{\DU}{\Delta U}

\newcommand{\Ga}{\mbox{Ga}}
\newcommand{\IGa}{\mbox{InvGa}}

\newcommand{\Weib}{\mbox{Weib}}
\newcommand{\Exp}{\mbox{Exp}}

\newcommand{\nmax}{N}
\newcommand{\Uo}{U^o}

\begin{document}
\doublespacing

\title{A Decision-Theoretic Comparison of Treatments to Resolve Air
  Leaks After Lung Surgery Based on Nonparametric Modeling}

\author{Yanxun Xu \\
{\small Dept. of Statistics and Data Sciences, University of Texas at Austin, Austin, TX, U.S.A.}
\and
Peter F. Thall\\
{\small Dept. of Biostatistics, University of Texas, M.D. Anderson Cancer Center, Houston, TX, U.S.A. }
\and
 Peter M\"uller  \thanks{Address for Correspondence:  pmueller@math.utexas.edu}\\
 {\small Dept. of Mathematics, University of Texas at Austin, Austin, TX, U.S.A. }
 \and Mehran J. Reza  \\
 {\small Dept. of Thoracic Surgery,  The University of Texas M.D. Anderson Cancer Center, Houston, TX, U.S.A. }
 }

\begin{center}
{\large \bf A Decision-Theoretic Comparison of Treatments to Resolve Air
  Leaks After Lung Surgery Based on Nonparametric Modeling}
\medskip

Yanxun Xu\footnote{
  Dept. of Statistics and Data Sciences, University of Texas at Austin}
 \footnote{Dept. of Applied Mathematics and Statistics, Johns Hopkins University},
Peter F. Thall\footnote{
  Dept. of Biostatistics, University of Texas, M.D. Anderson Cancer Center},
Peter M\"uller\footnote{
   Dept. of Mathematics, University of Texas at Austin,
     pmueller@math.utexas.edu},
and
Mehran J. Reza\footnote{
  Dept. of Thoracic Surgery,  The University of Texas M.D. Anderson Cancer Center}
\end{center}


\begin{abstract}
We propose a Bayesian nonparametric utility-based group sequential
design
for a randomized clinical trial to compare a gel sealant to
standard care for resolving air leaks after pulmonary resection.
Clinically, resolving air leaks in the days soon after surgery is
highly important, since longer resolution time produces undesirable
complications that require extended hospitalization.  The problem of
comparing treatments is complicated by the fact that the resolution
time distributions are skewed and multi-modal, so using means is
misleading.  We address these challenges by assuming Bayesian
nonparametric probability models for the resolution time distributions
and basing the comparative test on weighted means. The weights are
elicited as clinical utilities of the resolution times. The proposed
design uses posterior expected utilities as group sequential test
criteria.  The procedure's frequentist properties are studied by
extensive simulations.


\end{abstract}

\section{Introduction}
\subsection{The motivating clinical trial}
Intraoperative air leaks (IALs) occur in 48 to 75\% of patients after
pulmonary resection \citep{serra2005surgical}.  Despite the routine
use of intraoperative sutures and stapling devices, IALs remain a
significant problem in the practice of thoracic surgery. IALs
that persist beyond the immediate postoperative period of five days
may result in longer chest tube drainage, greater postoperative pain,
increased risk of infection, empyema, thromboemboli, and increased
length of hospitalization \citep{merritt2010evidence,
singhal2010management}.
Air leaks are a particularly severe problem in patients with
emphysematous lungs or who have undergone extensive visceral pleural
denuding procedures, such as pleurectomy decortication.  This is a
surgical procedure in which the lining surrounding one lung first is
removed (pleurectomy), and then any tumor masses that are growing
inside the chest cavity are removed (decortication).
In addition to the noted
risks to the patient, the economic impact of a prolonged air leak is
significant, primarily due to increased hospital stay.  Because the standard
procedure of suturing visible leaks and using staple reinforcement gives
unpredictable results, an alternative technique to control IALs is the
use of liquid sealants, which are thick fluids instilled in the areas
of leaks.  Progel (Neomend, Inc., Irvine, CA) is a polymeric
biodegradable hydrogel sealant, that currently is the only FDA
approved sealant to control IALs during pulmonary resection
\citep{kobayashi2001vivo}.

Despite FDA approval, the true benefit of Progel in reducing the rate of occurrence
or duration of IALs in lung resection patients has not been
established, and therefore it is not used routinely.
Researchers have conducted two studies
comparing Progel (treatment group) with standard care (control group)
to demonstrate the safety and efficacy of Progel
\citep{allen2004prospective, klijian2012novel}. Because the study of
\cite{allen2004prospective} varied the application of Progel based on
the size of the air bubbles seen in each patient, and the precise
methodology of how this was done was not explained in sufficient
detail to enable replication, the results of this trial are of limited
use for a general comparison of Progel to standard care. The study of
\cite{klijian2012novel} was retrospective and not randomized.  Given
these limitations of existing data, the desire to obtain a prospective
randomized comparison of Progel to standard care motivated the
clinical trial described in this paper.
The trial has passed IRB (internal review board)
approval and is scheduled to start accrual at
The University of Texas M.D. Anderson 
Cancer Center.

\subsection{Modeling considerations}
\begin{figure}
\centering
  \includegraphics[scale=0.45]{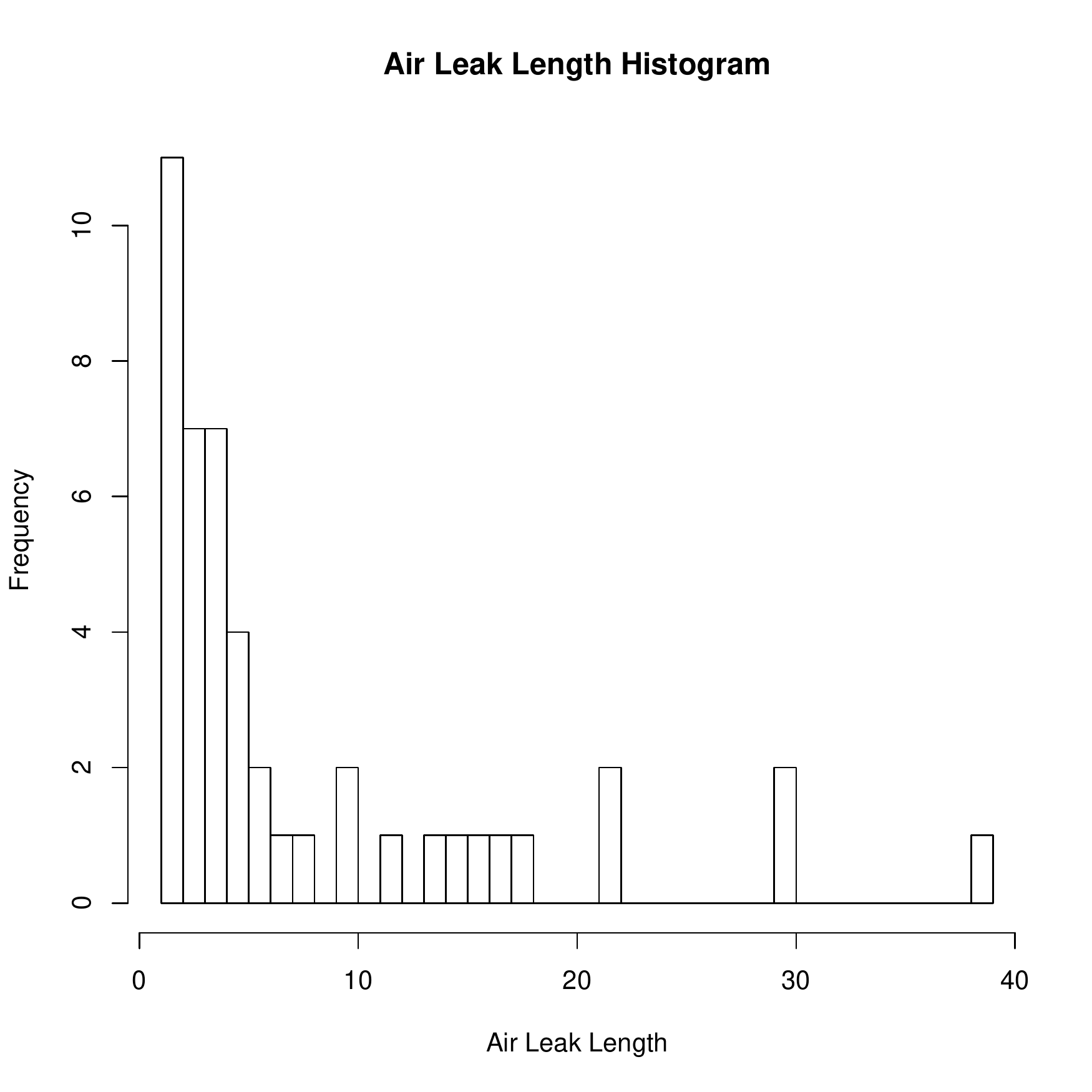}
\caption{Histogram of times to resolve lung air leaks, from the historical dataset. }
\label{fig:history}
\end{figure}

Denote by $T$ the days to resolve  IALs, allowing the possibility that an
air leak may not develop, represented by $T=0$.
\cite{allen2004prospective} and \cite{klijian2012novel} compared
$\mu_0$ and $\mu_1$, the means of $T$ in a control and treatment
group, respectively, using a standard $t$-test, and concluded that Progel
was superior to standard care in reducing IALs.
Figure \ref{fig:history}
plots the histogram of $T$ obtained from
non-randomized
historical data in the clinical database of
the Department of Thoracic and Cardiovascular Surgery at M.D. Anderson
Cancer Center.
The histogram suggests that a standard parametric model is
inappropriate to describe air leak resolution time distributions. For
example, a normal or log normal distribution would fail to
allow for the observed multi-modality and late resolution times.
Moreover, some patients treated with Progel after resection may be
free of air leaks immediately following surgery, corresponding to a
positive probability mass at $T=0$. 

Let $G_1$ denote the distribution of $T$ in the treatment (Progel) group
and $G_0$ the distribution of $T$ with the control (standard care).  We
will represent each $G_j$, $j=0,1$ as a mixture of a point mass
$\delta_0$ at $0$ and a hypothesized distribution $M_j$ for non-zero
resolution times, with $M_1$ a left-shifted version of $M_0$
to formalize the assumption that, stochastically, IAL  resolution times with
Progel are no longer
than with standard care.
This order constraint is motivated by several medical 
considerations:
  Progel is inert, and thus it cannot react chemically with the
patient's lung tissue, is not a potential source of infection, and
does not slow down the healing process.  Moreover, Progel cannot make
an air leak worse because it does not contribute to air leak
formation.  These considerations motivate {\it a priori} stochastic
ordering of $G_1$ and $G_0,$ which effectively says that, in terms of
time to resolve an IAL, Progel may be better than standard care, but
it cannot be worse. 
Nevertheless, for comparison we later report also inference under
an otherwise equivalent model without the stochastic ordering 
constraint. 

An important
consideration in developing a trial design is that the use of an
expected value 
as the target for a comparative test is inappropriate and inadequate,  both
because the historical distribution is skewed with a long right tail,
and because
a change in the early days after surgery is clinically more
important than a comparable change in later days.  Also, a standard test
of $\mu_1=\mu_0$ versus $\mu_1 < \mu_0$ would require an
impractically large sample size to achieve  any reasonable  power.
These complications are the principal reasons why designing a randomized trial to compare Progel to standard care is non-trivial, and why the
use of Progel has not been widely accepted among surgeons who
perform pulmonary resections.

The desire to obtain reliable confirmatory evidence to evaluate the
comparative benefit of Progel motivates the randomized trial described in this paper.  The goal of the trial is to assess the extent to which Progel is superior to standard care. The comparison also allows for the possibility no difference. 


\subsection{Stochastic ordering and  Bayesian nonparametric
  priors}
The time until resolution of air leaks for patients treated with
Progel is  {\it a priori}  expected to be shorter than under
standard care.
This introduces a stochastic ordering constraint on $G_0$ and $G_1$.
Formally, a distribution {\it $G_1$ is  stochastically smaller
than $G_0$},
denoted by $G_1 \preceq G_0$, if the
corresponding cumulative distribution functions satisfy $F_1(t)\geq
F_0(t)$ for all $t$.  \cite{lehmann2006testing} and
\cite{randles1979introduction} have modeled stochastic ordering
parametrically. Although straightforward, these approaches are limited
by the requirement that a parametric family must be specified.

To compare distributions of
air leak resolution times,  detailed
features (e.g., skewed or multi-modal) of the distributions are important, leading us to consider a
Bayesian nonparametric (BNP) approach. Importantly, uncertainties about the inference on these details are critical, as posterior probabilities about comparisons drive the decisions about sequential continuation and the terminal decision. Such descriptions of uncertainties are best considered in the framework of a probability model on the unknown distributions, as they are implemented in BNP models.

Formally, BNP refers to prior models for infinite dimensional unknown
quantities. Inference for random distributions, like $G_0$ and $G_1$ here, is a
typical example. A common feature of BNP models is their large support,
which allows one to approximate essentially arbitrary distributions \citep{ishwaran2001gibbs}.
For the proposed design, we use a model based on the Dirichlet process
(DP) prior \citep{ferguson1973bayesian}, which is
 by far the most commonly used BNP
model for a random distribution.
\cite{maceachern:99} introduced the dependent DP (DDP), which
extended the DP to a probability model for a family
$\{G_x,\; x \in X\}$ of random probability measures, indexed by some
covariate $x$.
The special case of a finite family, like $\{G_0,G_1\}$ in our
application, was discussed in \cite{deiorio&al:04}.
Several authors have considered BNP models for stochastically ordered
distributions.
\cite{gelfand2001nonparametric} started with two DP random probability
measures $G_0$ and $G_1$, and used the
product of the corresponding cumulative distribution functions
to define a pair of stochastically ordered random probability
measures.
A general methodology for stochastic ordering by considering
probability measures constrained to a convex set was proposed by
\cite{hoff2003bayesian}.
Finally, \cite{dunson2008bayesian} incorporated
stochastic ordering constraints in the DDP prior.
In this paper, we use
a simple implementation of this finite DDP model with order constraint
as the prior probability model for the  distributions $G_1$ and $G_0$
of leak resolution times under Progel and control.
Details are discussed in Section \ref{sec:model}.
For more extensive reviews of BNP methods, see, for example,
\cite{hjort;holmes;mueller;walker;2010}.

Based on the proposed BNP model, we define a utility-based
decision criterion to develop a clinical trial design.
To our knowledge, there is no literature 
on using BNP 
stochastic ordering models in construction of clinical
trial designs.  The main novelties in the proposed approach are the
successful use of utilities in a small scale clinical trial, a
convincing case for the need of a full probabilistic description of
uncertainties on random probability measures, and a simple and
practicable construction of a BNP prior on stochastically ordered
random probability measures with point masses.  
The trial will be
conducted in The University of Texas M.D. Anderson
Cancer Center, with a co-author of this paper (RM) its Principal Investigator.

Important practical advantages of the proposed approach are that
it allows meaningful borrowing of information from historical data (by
centering the BNP model), borrowing across treatments (by constructing
correlated priors on $G_0$ and $G_1$), and the exploitation of
stochastic ordering constraints, if warranted and approved in IRB
reviews. The use of utility weighting for the outcomes,
as  in this application, is  particularly natural under a BNP
model because it 
allows inference about all aspects of the event time distribution,
without constraint to parametric families. Together, these features
allow the investigators to plan a much smaller sample size than what
would be required  by a conventional trial design. For example, based
on the historical mean of 8 days and standard deviation 8.76, a two
sample one-sided 0.05-level $t$-test with power 0.80 to detect a 25\%
drop in the mean, from 8 to 6 days, would require a sample of $n =
476$ patients. This is impossible for this single-institution
trial. Given the realistic maximum target accrual of 48 patients, the
question is whether a design can be constructed that has reasonably
high power to conclude that Progel is superior to standard care under
clinically meaningful alternatives. We will show that the proposed
design, based on  differences in mean utilities evaluated from the
posterior under the BNP model, has very desirable operating
characteristics with $n = 48$ patients. The impact is the opportunity
to establish what is expected to be a greatly superior treatment
option for patients, with reasonable cost and effort. 


\section{Utilities and Trial Design}
\label{sec:trial}
The primary outcome is $T$, the time (in days) to resolve an air leak in the
lungs following surgery,  and we define
$Y=\log(T+1)$.
The possibility that an air leak may not develop is represented by $T=Y=0$.
The mean, or any other single measure of central tendency of $Y$,
is not an appropriate summary for treatment comparisons.
Instead, we take a utility-based approach. Utility-based decision criteria have been used recently
in clinical trial designs \citep{thall2012adaptive, lee2014bayesian}.
We use utilities to weight the importance of air leak
resolution times after surgery.
For example, a difference of a few days in time to resolution of air leaks in
the days immediately after surgery is far more important than a
comparable difference in later days.
We performed a formal utility elicitation with our clinical collaborator RM.
The rationale of the utility elicitation includes:
1) the most desirable resolution time is $T=0$ (free of air leaks
immediately after surgery, although this ideal outcome is almost never seen with standard care);
2)  early ($1\leq T\leq 5$) resolution of air leaks
is very desirable and
therefore the interval [1, 5] received a relatively high utility;
3) the utilities drop off steeply for later resolution times ($T>5$).
These considerations are both medical and economic, and they motivated  the elicited utilities $u(Y)$ for $Y=\log(T+1)$ in Table
\ref{table:utility}.
\begin{table}
  \centering
  \begin{tabular}{ c | ccccccccc }
    \hline
    $T$ (days) & 0 & 5 & 10 & 15 & 20 & 25 & 30 & 35 & $\geq 40$ \\ \hline
    Utility & 100 & 50 & 10 & 6 & 5 & 4 &3 &2 &0 \\
    \hline
  \end{tabular}
  \caption{Elicited utility  $u(T)$ for $T$ = days to resolve intra-operative air leak. }
  \label{table:utility}
\end{table}
 In the table,  the numerical utility 50 assigned to the outcome
$T=5$ days corresponds to the subjective assessment of RM that this
comparatively favorable outcome is half as desirable as the ideal
outcome of having  no air leak at all. Similarly, the utility for
$T = 10$ days reflects that this outcome, which involves a long hospital
stay and the complications described earlier, is 1/5 as desirable as
the outcome $T=5$ days.
We now are ready to define the expected utility for each group as
\begin{eqnarray}
  \Ubar_j= \int u(Y) G_j(dY), \ j=0, 1,
  \label{eq:uti}
\end{eqnarray}
where $G_j$ is a sampling model for the outcome in treatment group
$j$.
This expectation is over the distribution of the outcome $T$, and is conditional on the unknown distribution $G_j$.
We do not need to make any specific assumptions about $G_j$ yet,
except for the existence of such a distribution.


Based on the probability model and the utilities of Table
\ref{table:utility}, we now define a  design for the
Progel trial.
There are two types of decisions to be considered. At each interim test in the group sequential procedure,
we make a stopping decision $d_i \in \{0,1\}$
to stop ($d_i=0$) or continue ($d_i=1$).
If we reach a predetermined maximum sample size, $\nmax,$ we set $d_{\nmax}=0$ by
definition. Upon stopping, a terminal decision $a \in \{0,1\}$
reports the final recommendation, with $a=1$ denoting a recommendation
for Progel  and $a=0$ for standard care.
A decision-theoretic optimal solution would require backward induction \citep{bellman1957dynamic}
to solve the full sequential decision problem.
We stop short of carrying out this computationally prohibitive solution.
Instead, we propose to conduct the trial as follows.

\paragraph*{Sequential stopping rule.}
Patients are enrolled
in the trial sequentially in cohorts of size $m=16$ until a maximum of
$\nmax=48$ patients is reached or early stopping is indicated.
All patients are randomized equally  to control group and treatment
group, with the restriction of perfect balance after each cohort,
when the  continuation decisions $d_i$ are made.

Denote $\bY_{n} =\{Y_{ji},\; i=1,\ldots, n/2,\ j=0, 1\}$, the observed data for
the first $n$ treated patients, that is $n/2$ patients in each group
under the restricted equal randomization
 (rounding $n/2$ for odd $n$). 
The proposed decision criterion is the posterior probability
\begin{equation}
   \eta(\epsilon_U, \bY_n)= p(\Ubar_1 > \Ubar_0 +
   \epsilon_U \mid \bY_n),
\label{eq:u}
\end{equation}
where $\epsilon_U\geq 0$ is a minimum clinically meaningful difference
in expected utility.
Because the sequential rule makes multiple decisions, as with any
group sequential procedure the decision boundaries must be calibrated
to control the design's overall false positive error rate.  This is
similar to the use of so-called alpha-spending functions in
conventional frequentist group sequential designs. 
Like other frequentist summaries, false positive error rate
(type-I error) is a probability under an assumed truth, with respect
to repeated simulations of the entire trial. 
In the context of clinical trial designs such summaries
under repeated simulations are also known as (frequentist) operating
characteristics (OCs). 
Because evaluating
the design's OCs analytically is far too
complex, we do this by repeated computer simulations of the design,
under an array of different possible scenarios. This follows routine
practice in evaluating the behavior of sequentially adaptive clinical
trial designs.  In the present setting, the OCs are the type I error,
mean sample size, and probabilities of different possible decisions
(correct decision, stop due to futility, stop due to superiority).
Details are reported in Tables \ref{table:result} and S1.
\


After each cohort, we carry out Markov chain Monte Carlo posterior
simulation and evaluate the posterior estimates
$\etahat(\epsilon_U, \bY_n)$.
Let $\xi_U$ be an upper probability boundary for which the trial
will be terminated early and the treatment arm declared
superior if $\etahat(\epsilon_U, \bY_n)  \geq
\xi_U$. Similarly, let $\xi_L$ be a lower boundary for which the
trial will be terminated early due to futility, with the null hypothesis accepted,  if
$\etahat(\epsilon_U, \bY_n) \leq \xi_L$.
The bounds $\xi_L$ and $\xi_U$ are chosen by preliminary simulations
to obtain a design with desirable frequentist OCs. In Section
\ref{sec:trialsimu}, we will illustrate how one may calibrate these bounds.
In summary, the sequential stopping decision at any point of the trial
is: $d_n=1$ if $\xi_L < \etahat(\eps_U,\bY_n) < \xi_U$; $d_n=0$ otherwise.


\paragraph*{Terminal decision rule.}
Upon stopping, we record the terminal decision
$a=1$ if  \ $\etahat(\eps_U,\bY_n) > \frac12$ and
$a=0$ otherwise.
Assuming that $\xi_L < 0.5 < \xi_U$, the rule simply records whether
we stop due to crossing either the upper or lower bound, respectively.
If the trial reaches the maximum number of patients, $N=48$, the
terminal decision uses the threshold $\etahat > 0.5$ to determine
a recommendation for Progel.

\section{Probability Model}
\label{sec:model}
\subsection{Model and properties}
\label{sec:model1}
 We now construct a prior probability model for $G_j$, the
sampling model for $Y_{ij}$ for patients under control ($j=0$) and
Progel ($j=1$).
Because some patients may be free of air leaks immediately following
surgery, we allow a point mass at $Y_{ji}=0$ by defining
$G_j$, $j=0,1$, as mixtures
\begin{eqnarray}
   G_j &=& \nu_{j0} \delta_0 + \sum_{h=1}^{\infty}
    \nu_{jh} N(\theta_{jh},
   \sigma^2)
   = \nu_{j0} \delta_0 + (1-\nu_{j0}) \sum_{h=1}^{\infty}
    w_h N(\theta_{jh},
   \sigma^2) \nonumber\\
    &=& \nu_{j0} \delta_0 + (1-\nu_{j0}) M_j,  
\label{eq:model1}
\end{eqnarray}
where $\sum_{h=1}^{\infty} w_{h}=1$. Also, we impose a constraint
$\nu_{10} \geq \nu_{00}$ on the probabilities
$\nu_{10}$ and $\nu_{00}$,
and $M_1 \preceq M_0$,
formalizing the prior belief that patients are
more likely to be free of an air leak in the treatment group than in
the control group.
For  $M_j = \sum_{h=1}^\infty w_h N(\th_{jh}, \sig^2),$ $j=0,1$, we use a DDP prior
with common weights and dependent atoms.
The common weights $w_h$ have the DP stick-breaking prior,
$w_h = v_h\prod_{\ell < h} (1-v_\ell)$ with $v_h \sim
\mathrm{Beta}(1,\alpha)$.
The dependent prior on the atoms is constructed as follows, to ensure
$M_1 \preceq M_0$.
We assume $\bth_h = (\th_{0h},\th_{1h}) \sim \Ms$, where
$\Ms$ is a truncated multivariate normal base measure,
including a positive probability $\kappa$ for ties $\th_{0h}=\th_{1h}$:
\begin{eqnarray}
  \Ms(\bth_h)
  =
  N(\theta_{1h} \mid \mu_1, \sigma_1^2 )
  \left(\kappa     I(\theta_{0h}=\theta_{1h}) +
        (1-\kappa) N_{+}(\th_{0h} \mid  \th_{1h}, \tau^2) \right),
  \label{eq:prior}
\end{eqnarray}
where $N_{+}(x \mid m, V)$ refers to a truncated normal random
variable $x$ subject to $x \geq m$, and 
$\kappa=p(\theta_{0h}=\theta_{1h})$.   
For comparison we will also
consider inference under a variation of model \eqref{eq:prior} without
the order constraint, replacing the $N_{+}$ kernel by an unconstrained
normal $N(\theta_{1h},\tau^2)$. 

Denote $\tM_j = \sum_h w_h \delta_{\theta_{jh}}$, where $\delta_{\theta_{jh}}$ denotes a point mass at $\theta_{jh},\; j=0, 1$.
It is straightforward to show that $\tM_1 \preceq \tM_0$, which implies
$M_1 \preceq M_0$, and this in turn implies
$G_1 \preceq G_0$, as desired. \cite{barrientos2012support} study the
support properties of various DDP models. Applying Theorem 2 of
\cite{barrientos2012support},  it follows that 
the proposed model has full support over all pairs of stochastically ordered random probability measures.


 For reference we state the complete model,
\begin{eqnarray}
   Y_{ji} \mid \nu_{j0},\nu_{j1}, M_0,M_1 & \sim & G_j =
      \nu_{j0} \delta_0 + (1-\nu_{j0})
       \sum_{h=1}^{\infty} w_hN(Y_{ji}\mid \theta_{jh},
       \sigma^2)
                   \nonumber\\
      (\theta_{0h}, \theta_{1h}) \mid   \mu_1, \sig_1,
      \kappa,\tau & \sim &
      \Ms.
\label{eq:model3}
\end{eqnarray}
We complete the model specification with choices for the
hyperparameters
$\nu_{j0},\nu_{j1},\sig^2$, $\kappa, \mu_1, \sig_1, \tau$.
%
In the context of clinical trial design,  the
hyperparameters should not introduce inappropriately strong
information into the prior.
To ensure this, we provide the following guidelines.

We first standardize the data by subtracting the sample mean $\bar{Y}_1$ of the $Y_{1i}$'s
of the treatment group and scaling with the sample standard  deviation $s_1,$ mapping
$Y_{ji} \rightarrow$ $(Y_{ji}-\bar{Y}_1)/s_1.$
 This is done to
mitigate sensitivity to the measurement scale.
We fix $\mu_1=0$ and $\sig_1=1$, to reflect the standardization.
For $\sigma^2$, we assume
$p(1/\sigma^2) = \Ga(0.001, 0.001)$ to ensure that the prior is 
 not too informative,
where $\Ga(a, b)$ denotes a gamma distribution with mean $a/b$.
To allow for a wide range of shifts in
the response density, we specify
$p(1/\tau^2) = \Ga(0.5, 0.5)$.
This implies a Cauchy distribution for $\th_{0h}$,
which  often is used as a robust choice in parametric models.
To satisfy the constraint $\nu_{10}\geq \nu_{00}$, we  let
$\zeta_0=\nu_{00}$, $\zeta_1=\nu_{10}-\nu_{00}$, and  assume
$p(\zeta_0, \zeta_1)= \mathrm{Dirichlet}(0.1, 0.1, 0.1)$.
Finally, we assume $p(\kappa)= \mathrm{Beta}(1,1)$ and $p(\alpha) = \Ga(1,1)$.
The conjugacy of the implied normal on $\bth_h$ in (\ref{eq:prior})
and the normal kernel in (\ref{eq:model3}) greatly simplify posterior
inference. Any Markov chain Monte Carlo (MCMC) scheme for DP mixture
model as described, for example, in  \cite{neal2000markov},
can be applied.
In our implementation, we used 
 an implementation based on the 
finite DP \citep{ishwaran2001gibbs}, which truncates the
infinite sum in the DP mixture model after a finite number of
terms. We used $H=10$, following a recommendation based on Theorem
1 in \cite{ishwaran2002approximate} that gives tight bounds on
the approximation error, well below what is clinically relevant in this
application.
Details of the MCMC implementation are presented in Supplement A.



We carried out a preliminary simulation study to better understand the
nature and accuracy of
posterior inference under the proposed model for a reasonable sample
size. The simulation setup and results are summarized in Supplement B. The inference under
the proposed method incorporating the stochastic ordering constraint
performed well, indicating small bias even with moderate sample size.

\section{Trial Simulation Study}
\label{sec:trialsimu}

To assess average behavior of the proposed BNP trial design, we
performed an extensive simulation study under a variety of scenarios
that were constructed to  mimic the Progel trial. 
For the proposed stopping and decision
rules, we fixed the parameters as $\xi_U=0.9, \xi_L=0.05$,
based on preliminary studies (described later) and examining the OCs of the proposed BNP design. In all scenarios,
we set the maximum number of patients to be $N=48$, randomized  equally
between the control and treatment group, with  cohorts of 16
patients.  The smallest clinically meaningful improvement 
used to define the decision criterion $\eta(\epsilon_U, \bY_n)$
was determined by our clinical collaborator (RM)
 to be $\epsilon_U=18$, given the numerical utilities of IAL resolution times
 in Table 1.

We considered nine scenarios, and simulated 100 trials for each
scenario. The response outcomes $Y_{ji}$ were generated from the 
simulation truth $G^o_j$ shown in the last column 
of Table \ref{table:setup}.
 Other columns in the same table show 
the true utilities $\Uo_j = \int u(y)\, dG^o_j(y)$ for each arm and the
differences $\Uo_1-\Uo_0$.

 \begin{table}
 \caption{In each scenario, the models $G^o_j$ in the right column are the simulation truths.
   Here $\bar{\sigma}=0.3$ and $\Exp(\cdot),
   \Weib(\cdot,\cdot)$ denote an exponential distribution and a Weibull
   distribution, respectively.  $\Uo_j$ reports the expected utilities
   under the simulation truth $G^o_j$. The second column reports the
   true difference $\Uo_1-U^o_0$.}
\centering
\begin{tabular}{ccccc}
Scenario & $\Uo_1-\Uo_0$ & Group&  $\Uo_j$ & Simulation truth $G^o_j$\\
\hline
	& 			& Progel &- & Resample from historical data \\
\multirow{-2}{*}{1} 	& \multirow{-2}{*}{0} &Control& - &   Resample from historical data  \\

\hline
	& 			& Progel &23.44 & $0.1\delta_0 + 0.63 N(2,  \bar{\sigma}^2) + 0.27 N(3, \bar{\sigma}^2) $ \\
\multirow{-2}{*}{2} 	& \multirow{-2}{*}{0} &Control& 23.44 &   $0.1\delta_0 + 0.63 N(2,  \bar{\sigma}^2) + 0.27 N(3, \bar{\sigma}^2) $  \\
\hline
	& 			& Progel &57.25 & $0.3\delta_0 + 0.49 N(1,  \bar{\sigma}^2) + 0.14 N(2, \bar{\sigma}^2) + 0.07 N(3.5,  \bar{\sigma}^2)$ \\
\multirow{-2}{*}{3} 	& \multirow{-2}{*}{41.64} &Control& 15.61 &   $0.1\delta_0 + 0.63 N(2.5,  \bar{\sigma}^2) + 0.18 N(3,  \bar{\sigma}^2) + 0.09 N(4.5,  \bar{\sigma}^2)$  \\
\hline
	& 			                                           & Progel &48.94 & $0.2\delta_0 + 0.56 N(1.5,  \bar{\sigma}^2) + 0.24 N(2, \bar{\sigma}^2) $  \\
\multirow{-2}{*}{4} 	& \multirow{-2}{*}{19.15} &Control& 29.79 &   $0.1\delta_0 + 0.63 N(1.8,  \bar{\sigma}^2) + 0.27 N(3, \bar{\sigma}^2) $  \\
\hline
	& 			                                           & Progel &64.82 & $0.4\delta_0 + 0.48 N(1,  \bar{\sigma}^2) + 0.12 N(2.5, \bar{\sigma}^2) $  \\
\multirow{-2}{*}{5} 	& \multirow{-2}{*}{29.68} &Control& 35.14 &   $0.1\delta_0 + 0.54 N(1.5,  \bar{\sigma}^2) + 0.18 N(2.5, \bar{\sigma}^2) + 0.18 N(3.5, \bar{\sigma}^2) $  \\
\hline
	& 			                                           & Progel &60.80 & $0.4\delta_0 + 0.36 N(1,  \bar{\sigma}^2) + 0.12 N(2, \bar{\sigma}^2) + 0.12 N(3, \bar{\sigma}^2)$  \\
\multirow{-2}{*}{6} 	& \multirow{-2}{*}{34.15} &Control& 26.65 &   $0.1\delta_0 + 0.36 N(1.5,  \bar{\sigma}^2) + 0.54 N(3.5, \bar{\sigma}^2) $  \\
\hline
	& 			& Progel &55.33 & $0.3\delta_0 + 0.3 Exp(1) + 0.4Exp(0.5)$ \\
\multirow{-2}{*}{7} 	& \multirow{-2}{*}{43.47} &Control& 11.86 &   $0.1\delta_0 + 0.4 N(3,  0.2^2) + 0.5 N(4,  0.2^2) $  \\
\hline
	& 			& Progel &45.08 & $0.2\delta_0 + 0.4 \Weib(1,2) + 0.4 \Weib(0.7, 2)$ \\
\multirow{-2}{*}{8} 	& \multirow{-2}{*}{8.13} &Control& 36.95 &   $0.2\delta_0 + 0.5 N(1.8,  0.2^2) + 0.3 N(2.5,  0.3^2)$  \\
\hline
	& 			& Progel &55.33 & $0.3\delta_0 + 0.3 Exp(1) + 0.4Exp(0.5)$ \\
\multirow{-2}{*}{9} 	& \multirow{-2}{*}{10.25} &Control& 45.08 &   $0.2\delta_0 + 0.4 \Weib(1,2) + 0.4 \Weib(0.7, 2)$  \\
\hline
\end{tabular}
\label{table:setup}
\end{table}

To calculate type I error and power, we define the null hypothesis $H_0: G_1=G_0.$ 
Under the proposed design, the test
rejects $H_0$ in favor of Progel if $\etahat(\epsilon_U, \bY_n)\geq \xi_U$
interimly with early stopping at $n$=16 or 32, and if
$\etahat(\epsilon_U, \bY_N)\geq 0.5$  for the terminal rule at 
$N=48$.
Similarly, the test fails to reject $H_0$ if
$\etahat(\epsilon_U, \bY_n)\leq \xi_L$ ($n=16, 32$), with early stopping for futility,  and if
$\etahat(\epsilon_U, \bY_N)< 0.5$  at $N=48$.

We fixed the hyperparameters as described earlier in Section
\ref{sec:model} and fit the proposed BNP model (\ref{eq:model3}) to
each simulated data set.   Table \ref{table:result} summarizes the
OCs of the proposed BNP utility-based design for
nine scenarios. The OCs  include the average number of patients treated,
type I error, the probabilities of making the correct decision (PCD),
 stopping  the trial early due to either superiority, Pr(EarS),
or futility, Pr(EarF) and, in a final analysis without early stopping,
declaring superiority, Pr(FinS), or futility Pr(FinS).  
For comparison we also implemented inference under a variation of
the model without the stochastic ordering constraint, that is, model
\eqref{eq:prior} with an unconstrained normal $N(\th_{0h} \mid
\th_{1h},\tau^2)$ replacing the truncated normal in \eqref{eq:prior}.
Scenarios 1a and 2a show summaries of inference under this
unconstrained version of the model, using the same simulation truths
as in scenarios 1 and 2.

Details of the simulation results are discussed in
Supplement C.
\underline{Scenarios 1 and 2} are null scenarios;
in \underline{Scenario 3} we assumed a large treatment effect of
$\Uo_1-\Uo_0=41.6$, far beyond $\eps_U=18$;
\underline{Scenario 4} has a small treatment effect of
$\Uo_1-\Uo_0=19.2$, barely beyond $\eps_U$;
under \underline{Scenarios 5 and 6} we assumed a moderate treatment
effect; and the last three \underline{Scenarios 7, 8, and 9} have simulation truths different from the assumed mixture of normal
distributions.

For reference we also evaluated summaries related to estimation. \ech
Denote the true utility difference under the simulation truth by
$\DU^\star = \Uo_1-\Uo_0$, and $\DU = \Ubar_1- \Ubar_0$. For each
scenario, we computed the estimation bias $E\{\DU - \DU^\star\}$ and
root mean squared error (RMSE) $ \sqrt{E\{(\DU-\DU^\star)^2\}}$, where
the expectation is over repeated simulations under each scenario. The
results are given in Table S1.

The OCs under all nine scenarios
are given in Table \ref{table:result}, and
show a favorable evaluation of the proposed design.
The results under scenarios 1a and 2a, compared to scenarios 1 and 2,
show that the design's frequentist OCs  
appear to be  robust with respect to the inclusion or not of 
the constraint $G_0 < G_1$ in the prior. The simulation truths in scenarios 7, 8, and 9 are different from the assumed mixture of normal distributions with equal variance. The results under these scenarios  demonstrate the flexibility of BNP mixture models with a common variance parameter.  
In summary, inferences under the proposed BNP model and trial
monitoring rules exhibit desirable OCs across
all nine scenarios.

\begin{table}
  \caption{Trial simulation results. MSS = mean sample
    size,  TIE = type I error,  PCD = probability of making the
    correct decision, $Pr$(EarS) =  probability of
    stopping early due to superiority,  $Pr$(EarF) =  probability
    of stopping early due to futility, $Pr$(FinS) =  probability of
     declaring superiority in a final analysis without early stopping, and  $Pr$(FinF) =  probability of declaring
     futility in a final analysis without early stopping.
    All probabilities are computed by repeated simulations.}
  \centering
  {\small
\begin{tabular}{ccccc|cccc}
Scenario & $\Uo_1-\Uo_0$ & MSS& TIE & PCD & $Pr$(EarS) &  $Pr$(FinS) & $Pr$(EarF) &$Pr$(FinF)\\
\hline
1& 0 & 16.80& 0.00& 1.00 &0.00& 0.00 & 1.00&0.00\\
\hline
1a& 0 & 16.32& 0.00& 1.00 &0.00& 0.00 & 1.00&0.00\\
\hline
2& 0 & 28.80& 0.02& 0.98 &0.01& 0.01 & 0.64& 0.34\\
\hline
2a& 0 & 28.00& 0.01& 0.99 &0.00& 0.01 & 0.69&0.30\\
\hline
3& 41.64 &29.12& - & 1.00 & 0.81& 0.18 &0.01 & 0.00\\
\hline
4& 19.15 & 40.16& -& 0.63 & 0.15& 0.48 &0.15 & 0.22\\
\hline
5& 29.68 & 31.68& -&0.93 & 0.60& 0.33 &0.04 & 0.03\\
\hline
6& 34.15 & 29.92& -& 0.94& 0.65& 0.29 &0.05 &0.01\\
\hline
7&43.47  & 28.64& -& 0.96& 0.77& 0.19 &0.04 & 0.00\\
\hline
8&8.13 & 34.08& -& 0.79& 0.04& 0.17 &0.45 & 0.34\\
\hline
9&10.25 & 34.88& -& 0.74& 0.10& 0.13 &0.34 & 0.43\\
\hline
\end{tabular}
}
\label{table:result}
\end{table}

\paragraph*{Parametric models and sensitivity analyses.}
For comparison, we implemented alternative inference under
a parametric model assuming a zero-enriched Weibull distribution, that
is, a mixture of point mass at 0 and Weibull distribution.
We assumed $T_{ij} \iid G^W_j, i=1, \dots, n_j$ for groups $j=0$ and
$j=1$, using
$
  G^W_j=\pi_{j}\delta_0+(1-\pi_j)\Weib(\lambda_{1j}, \lambda_{2j}).
$
We completed the model with a prior
$p(\pi_j)=\mathrm{Beta}(0.1, 0.1)$ and a conjugate prior
$p(\lambda_{2j})=\IGa(b_{1j}, b_{2j})$.
The hyperparameters $b_{1j}$ and $b_{2j}$ were determined by
matching the prior mean of $\lambda_{2j}$ with a
maximum likelihood estimate and assuming a prior variance of
10.
Finally, for $\lambda_{1j}$ there is no conjugate prior. We followed
\cite{fink1997compendium} by assuming
$p(\lambda_{1j})\propto \lambda_{1j}^{a_{1j}}\exp (-a_{2j}\lambda_{1j}-\frac{a_{3j}^{\lambda_{1j}}}{\lambda_{2j}})$,
with $a_{1j}=1, a_{2j}=\log(\prod_{i=1}^{n_j}T_{ij})+2$, and $a_{3j}=2$, $j=0, 1$.

Table \ref{table:compare} shows the OCs
comparing the inferences under the proposed model  versus  the zero-enriched
parametric Weibull model in some (arbitrarily) selected scenarios.
The
proposed BNP model with stochastic ordering compares quite favorably,
with much larger probabilities of making a correct decision and 
correctly stopping early for superiority.

Finally, we carried out  an alternative  analysis to understand
how much the results might change if different utilities $u(t)$ were
elicited.
Table S2 in the supplement presents the results of a sensitive
analysis using different utilities.
In summary, while the actual decisions naturally change, the
 frequentist OCs change only
slightly. 
Different decisions are desirable, under different utilities
that reflect different clinical preferences. 

A final set of simulations explored robustness with respect to the
decision boundaries $\xi_U$ and $\xi_L$ for the continuation
decision.
Table S3 summarizes OCs
under Scenarios 2, 3, and 4.
Again, while some summaries, like the probability of early
stopping for futility, change in the expected direction, the nature
of the overall comparison across scenarios remains unchanged under
different criteria.

 \begin{table}
 \caption{Comparisons in selected scenarios under the proposed BNP
   model with stochastic ordering (BNPSO) versus an alternative
   parametric model with a zero-enriched Weibull
   (Z-Weib). }
\centering
  {\small
\begin{tabular}{cccccccccc}
Scenario & True Diff & Group&  MSS&PCD & $Pr$(EarS) &  $Pr$(FinS) & $Pr$(EarF) &$Pr$(FinF)\\
\hline
	& 			& BNPSO &29.12 &1.00&0.81&0.18&0.01&0.00  \\
\multirow{-2}{*}{3} 	& \multirow{-2}{*}{41.64} &Z-Weib& 44.32& 0.65&0.10&0.52&0.04&0.34    \\
\hline
	& 			& BNPSO &40.16 &0.63 & 0.15& 0.48 &0.15 & 0.22  \\
\multirow{-2}{*}{4} 	& \multirow{-2}{*}{19.15} &Z-Weib& 42.40&0.17&0.03&0.14&0.18& 0.65   \\
\hline
	& 			& BNPSO &29.92&0.94& 0.65& 0.29 &0.05 &0.01  \\
\multirow{-2}{*}{6} 	& \multirow{-2}{*}{34.15} &Z-Weib&40.00& 0.85&0.28&0.57&0.02&0.13    \\
\hline
	& 			& BNPSO &28.64 &0.96& 0.77& 0.19 &0.04 & 0.00  \\
\multirow{-2}{*}{7} 	& \multirow{-2}{*}{43.47} &Z-Weib&41.92& 0.87&0.20&0.65&0.04& 0.11   \\
\hline
	& 			& BNPSO &34.08 &0.79& 0.04& 0.17 &0.45 & 0.34  \\
\multirow{-2}{*}{8} 	& \multirow{-2}{*}{8.13} &Z-Weib&45.28& 0.42&0.06&0.50 &0.05&0.39   \\
\hline
\end{tabular}
}
\label{table:compare}
\end{table}

%
%

\section{Conclusions and Discussion}
\label{sec:dis}
We developed a Bayesian nonparametric (BNP) utility-based group sequential design to compare Progel with standard care in resolving air leaks after lung surgery. In this setting, standard statistical tests or parametric models are not appropriate for trial designs or to describe air leak resolution time distributions. We solved the problem by developing a BNP model with a stochastic ordering constraint and proposing a trial design based on  expected utility, computed from elicited utility values.
The model assessment and trial simulation studies show unbiased results and desirable OCs.

Beyond the application discussed in this paper, the proposed BNP utility-based method can be extended to many other contexts. For example, in applications that involve multiple groups, one may replace the truncated bivariate normal base measure $M^*$ in (\ref{eq:model3}) with a truncated multivariate normal distribution that incorporates the desired stochastic ordering constraints. Furthermore, the hypothesis testing framework discussed in Section \ref{sec:trialsimu}  can be  extended easily to testing equalities in multiple distributions that are  stochastically ordered.

Finally, we note that the BNP model could be replaced by a
  sufficiently flexible parametric model without any substantial change
  in the performance of the proposed design. For example, one could
  use a mixture of $H=5$ normals as the model. However, the computational
  effort for posterior simulation in any finite mixture of normal
  model is nothing less than in the proposed DDP model. We prefer the
  BNP model for reasons of conceptual clarity and, in principle,
  natural scaling to larger sample sizes and greater precision. 

\section*{Acknowledgement}

Peter M\"uller and Yanxun Xu's research  was partially supported by NIH/NCI grant RO1 CA 083932.  Peter Thall's research was supported by NCI grant R01 CA 83932.

\bibliographystyle{apalike}
\bibliography{progel}

\end{document}